\documentclass[doublecol]{epl2} 

\usepackage{graphicx}
\usepackage{dcolumn}
\usepackage{bm}
\usepackage{color}
\usepackage{appendix}
\usepackage{ulem}
\usepackage{xfrac}
\usepackage{amsmath}
\usepackage{amssymb}
\usepackage{ulem}

\newcommand{\eavg}[1] {\langle #1 \rangle}

\newcommand{\pd}[2] {\frac{\partial #1}{\partial #2}}

\newcommand{\vk}{von K\'{a}rm\'{a}n }

\title{Statistics of phase fluctuations of an acoustic wave propagating through a turbulent flow}

\author{G. Prabhudesai\inst{1} \and S. Perrard\inst{1,2} \and F. P\'{e}tr\'{e}lis\inst{1} \and S. Fauve\inst{1}}
\shortauthor{G. Prabhudesai \etal}

\institute{                    
  \inst{1} Laboratoire de Physique de l'\'{E}cole Normale Sup\'{e}rieure, CNRS, PSL Research University, Sorbonne Universit\'{e}, Universit\'{e} de Paris, F-75005 Paris, France \\
  \inst{2} PMMH, CNRS, ESPCI Paris, Université PSL, Sorbonne Université, Université Paris-Cité, 75005 Paris, France
}

\abstract{
We investigate the statistics of phase fluctuations of an acoustic wave propagating through a turbulent flow in line of sight (LOS) configuration. Experiments are performed on a closed \vk swirling flow whose boundaries are maintained at a constant temperature. In particular, we analyze the root mean square (RMS) and the power spectrum density (PSD) of phase fluctuations. A model is developed and analytical predictions obtained for these quantities using geometrical acoustics are shown to be in agreement with experimental observations. }

\begin{document}

\maketitle

\section{Introduction} 

Lighthill \cite{lighthill1952sound,lighthill1954sound} in two seminal articles first studied the problem of sound generated aerodynamically as intrinsic to an unsteady, inviscid and incompressible flow. This phenomenon of unsteady flow generating sound can be further extended to study the scattering of acoustic waves by flows. The earliest of studies on sound-flow interaction were presented by Rayleigh \cite{rayleigh1896theory} who treated the problem of refraction of sound waves by flows. Subsequent work by Obukhov \cite{obukhov1941scattering}, Blokhintzev \cite{blokhintzev1946propagation}, Kraichnan \cite{kraichnan1953scattering} and others led to further development in this field and the study of sound scattering due to velocity gradients in the flow. Fabrikant \cite{fabrikant1982,fabrikant1983} and Lund and Rojas \cite{lund1989ultrasound} related the scattered acoustic field with the vorticity field in the flow and gave compact formulae in the far-field limit under the approximations of Born and low Mach number of the flow. This theory was validated using experimental studies on both laminar and turbulent flows and used to characterize vorticity filaments (see for example \cite{gromov1982sound,baudet1991spectral,labbe1996ecoulements,dernoncourt1998experimental,berthet2001interaction,berthet2003,seifer2004flow,seifer2005spatial,hernandez2010}). Apart from studying the scattered acoustic field due to vorticity inhomogeneities in the flow, there is another effect which is equally if not more important due to its physical implications. As Tatarskii \cite{tatarski2016wave} notes, the scattering due to turbulent flow also results in parameter fluctuations of the incoming sound wave in the particular case of line-of-sight (LOS) propagation, i.e., when the scattering angle is zero. This effect is not restricted to only acoustic waves but can also be encountered in electromagnetic wave propagation and is a source of noise for optical ground based telescopes. To our knowledge, previous studies on the scattering of acoustic wave have focused on their spatial characteristics. In this letter, we show that the temporal characteristics of phase fluctuations of a LOS acoustic wave travelling through a turbulent flow can be related to the spatio-temporal characteristics of the flow. In particular, compact formulae are derived relating the RMS of phase fluctuations ($\Phi_{rms}$) to the characteristics of the turbulent flow, and, the frequency PSD of phase fluctuations ($E_{\Phi}$) to the spatio-temporal coherence of velocity fluctuations ($\mathcal{C}_{u}$). Generating a \vk turbulent swirling flow in air between counter-rotating disks, we experimentally validate the derived analytical results.

\begin{figure}[ht!]
	\centering
	\includegraphics[height=9cm,width=0.9\linewidth]{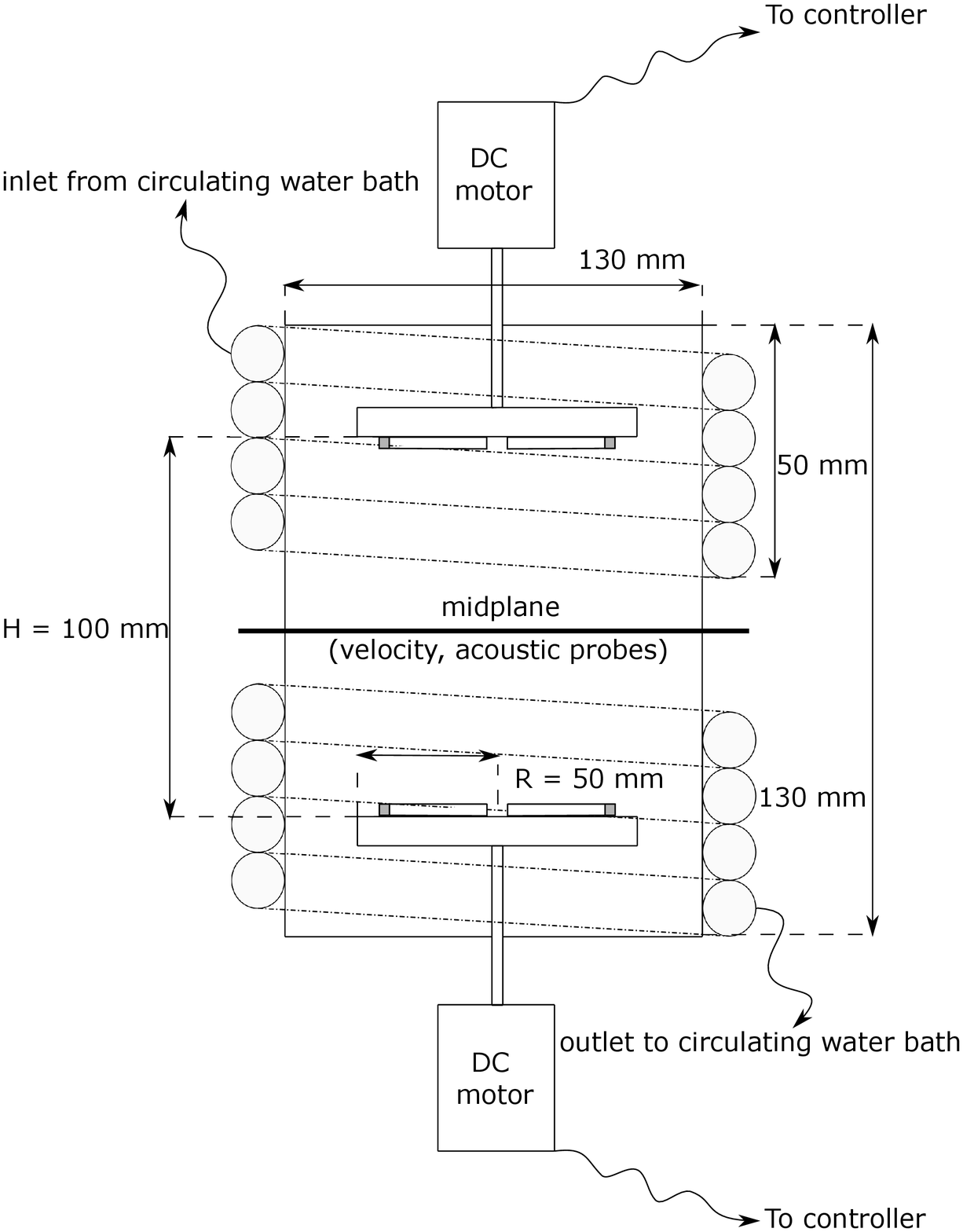}
	\includegraphics[height=6cm,width=0.9\linewidth]{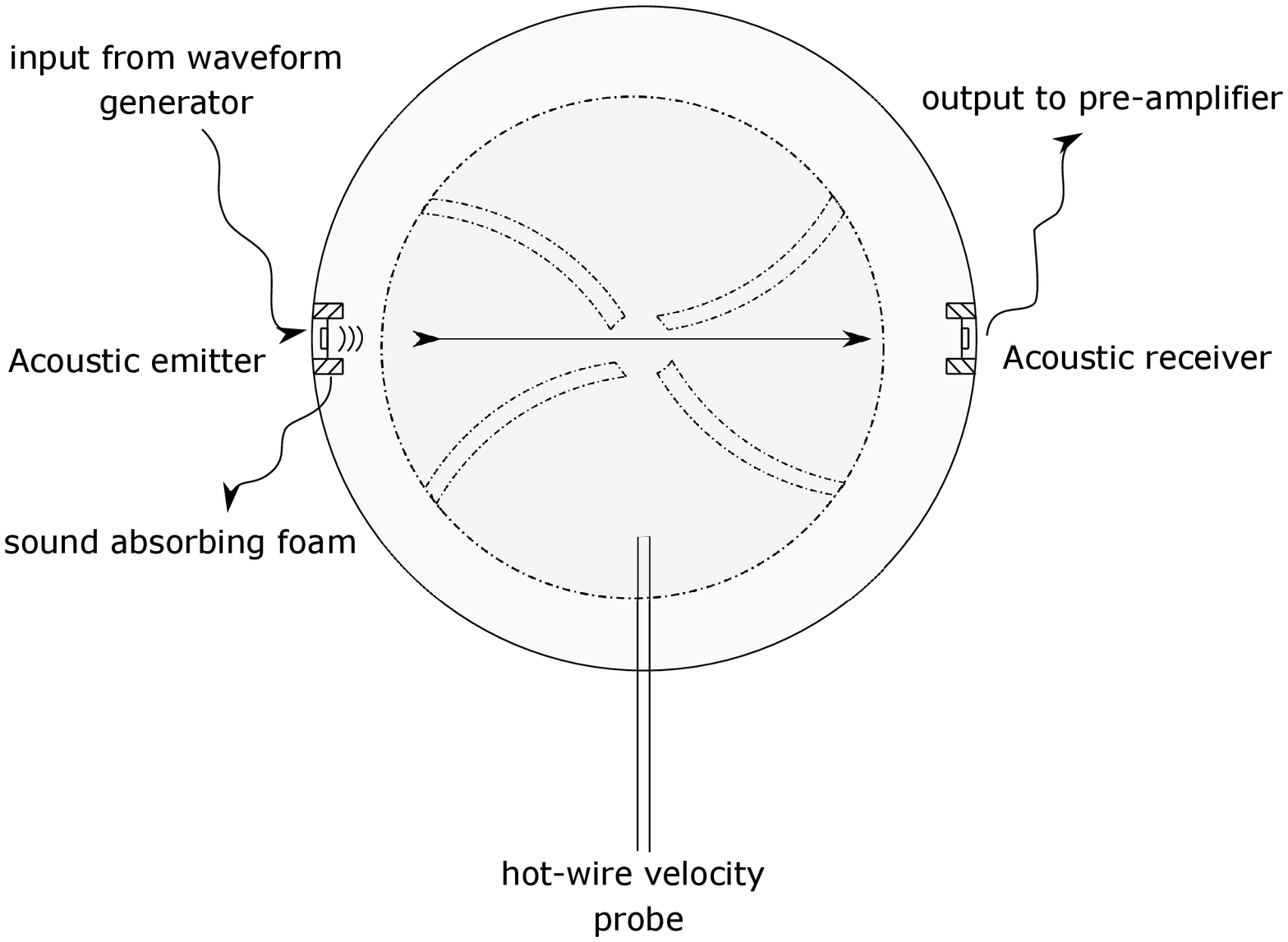}
	\caption{Sketch of the \vk flow configuration driven by two counter-rotating disks. Temperature of the boundaries is maintained constant using a circulating water bath. The velocity probe and acoustic transducers are all placed in the midplane.}
	\label{fig:1}
\end{figure}

\section{Theory}  

We consider an acoustic wave travelling through a turbulent flow of low Mach number under the geometrical acoustic limit $(k_{I}l_{0}/2\pi) \gg 1$. We have denoted the wavenumber of the incident acoustic wave by $k_{I}$ and the integral length scale by $l_{0}$, which corresponds to the correlation length of the turbulent velocity fluctuations. The geometrical acoustic limit implies that the wavelength of the acoustic wave is smaller than the length scale at which most of the energy in the turbulent flow is concentrated, the latter being same as $l_{0}$. The condition for geometrical acoustics can also be written in terms of frequency as $(f_{0}/f_{I}) \ll (u_{rms}/c)$, where $u_{rms}$ is the RMS of turbulent velocity fluctuations, $c$ is the speed of sound and $f_{0} = u_{rms}/l_{0}$ is the integral frequency. Using Rytov's method \cite{rytov1937diffraction}, we obtain the equation for fluctuations in its phase denoted by $\Phi$ (first Rytov approximation) \cite{tatarski2016wave},

\begin{align}
	\pd{\Phi(x,t)}{x} &= -k_{I}\Bigg(\frac{u_{x}(x,t)}{c}\Bigg) \,,
\end{align}

where $x$ is the direction of propagation of the acoustic wave, and $u_{x}$ is the $x$-component of turbulent velocity fluctuations. The above equation can be integrated over space to give the fluctuations in phase $\Phi_{L}(t)$ at a distance $L$ from the source,

\begin{align}
	\Phi_{L}(t) = -\Bigg(\frac{k_{I}}{c}\Bigg)\int_{0}^{L}\ u_{x}(x,t) \ dx \,,  
	\label{eqn:2}
\end{align}

where $\Phi_{L}(t) = \Phi(L,t)$. Using eqn.~\ref{eqn:2}, we can relate the statistical properties of $\Phi_{L}$ to that of $u_{x}$. First, $\eavg{\Phi_{L}} = 0$ since $\eavg{u_x} = 0$. The angle brackets denote the operation of averaging over time. Next, the second moment of $\Phi_{L}$ when the turbulent flow is homogeneous and isotropic is given by

\begin{align}
	\eavg{\Phi_{L}^{2}} &= \frac{4k_{I}^{2}u_{rms}^{2}}{c^{2}}\int_{0}^{L}\ dz\ \int_{0}^{z}\ \Gamma_{u}(r) \ dr \,,
	\label{eqn:3}
\end{align}

where the two point spatial correlation of velocity fluctuations for a homogeneous and isotropic flow is denoted by $\Gamma_{u}(r) = \eavg{u_{x}(x)u_{x}(x+r)}/u_{rms}^{2}$ \cite{tennekes1972first}. We consider a turbulent flow of large Reynolds number $Re = u_{rms}l_{0}/\nu \gg 1$ ($\nu$ is the kinematic viscosity of the fluid). 
We have measured in several experimental configurations \cite{prabhudesai2021fluctuations}, including the present one,  that the two point correlation decays exponentially for values of $r$ larger than a fraction of $l_0$ with a form $\Gamma_{u}(r) = \exp(-r/l_{0})$. On using this functional form for $\Gamma_u$ in eqn.~\ref{eqn:3} and taking square root, we obtain

\begin{align}
	\Phi_{rms} = \Bigg(\frac{2k_{I}u_{rms}}{c}\Bigg) \sqrt{l_{0}L} \,,
	\label{eqn:4}
\end{align}

where $\Phi_{rms} = \sqrt{\eavg{\Phi_{L}^{2}}}$. Similarly, on evaluating the temporal correlation of $\Phi_{L}$ using eqn.~\ref{eqn:2} and taking Fourier transform, we obtain an expression for its PSD, $E_{\Phi}(f) = \int_{-\infty}^{\infty} \eavg{\Phi_{L}(t)\Phi_{L}(t+\tau)} e^{2\pi \iota f \tau} \ d\tau$, which is

\begin{align}
	E_{\Phi}(f) = \Bigg(\frac{2k_{I}^{2}}{c^{2}}\Bigg)\Big(E_{u}(f)\Big)\int_{0}^{L}\ dz\ \int_{-z}^{z}\ dr\ \sqrt{\mathcal{C}_{u}(r,f)} \,,
	\label{eqn:5}
\end{align}

where $\mathcal{C}_{u}$ is the coherence and $E_{u}(f) = \int_{-\infty}^{\infty} \eavg{u_{x}(t)u_{x}(t+\tau)} e^{2\pi \iota f \tau} \ d\tau$ is the PSD of $u_{x}$ respectively. When the turbulent flow is homogeneous and isotropic, the coherence has the definition

\begin{align}
	\mathcal{C}_{u}(r,f) &= \frac{\Big|E_{u}(r,f)\Big|^{2}}{\Big(E_{u}(f)\Big)^{2}} \,,
	\label{eqn:6}
\end{align}

with $0 \leqslant \mathcal{C}_{u} \leqslant 1$. In eqn.~\ref{eqn:6}, $E_{u}(r,f) = \int_{-\infty}^{\infty} \eavg{u_{x}(x,t)u_{x}(x+r,t+\tau)} e^{2\pi \iota f \tau} \ d\tau$ is the cross PSD of $u_{x}$. While $E_{u}(f)$ is a real quantity, $E_{u}(r,f)$ is complex and $|\cdot|$ denotes its modulus. Recently, Prabhudesai \etal \cite{prabhudesai2021coherence} experimentally studied coherence of velocity fluctuations in turbulent flows and reported that it has the form

\begin{align}
	\mathcal{C}_{u}(r,f) = \exp\Bigg[-\frac{c_{1} r}{l_{0}}\Bigg(1 + \frac{c_{2} l_{0}f}{u_{rms}}\Bigg)\Bigg]  
	\label{eqn:7}
\end{align}

for $r/l_{0} \geqslant 0.27$. In this experiment, we measure  $c_{1} = 0.9$ and $c_{2} = 8$ which have the same order of magnitude as the values reported in \cite{prabhudesai2021coherence} for two different experimental configurations. On substituting this functional form in eqn.~\ref{eqn:5}, we obtain

\begin{align}
	E_{\Phi}(f) = \Bigg(\frac{8k_{I}^{2}l_{0}^{2}}{c_{1}^{2}c^{2}}\Bigg)\zeta(f)E_{u}(f) \,,
	\label{eqn:8}
\end{align}

where

\begin{align}
	\zeta(f) = \Bigg(&\frac{1}{1 + \frac{c_{2}l_{0}f}{u_{rms}}}\Bigg)\Bigg[ \Bigg(\frac{c_{1}L}{l_{0}}\Bigg) - \Bigg(\frac{2}{1 + \frac{c_{2}l_{0}f}{u_{rms}}}\Bigg) \nonumber \\ 
	&\Bigg( 1 - \exp\Bigg(-\frac{c_{1}L(1 + \frac{c_{2}l_{0}f}{u_{rms}})}{2l_{0}}\Bigg) \Bigg) \Bigg] \,.
	\label{eqn:9}
\end{align} 

The above analysis shows that when the velocity fluctuations at two points in space are perfectly coherent (i.e. when $	\mathcal{C}_{u} =1$), we have $E_{\Phi} \propto E_{u}$. The loss of coherence of velocity fluctuations results in the energy reduction of phase fluctuations, captured by the function $\zeta(f)$. Note that eqn.~\ref{eqn:4} (and eqn.~\ref{eqn:8}) also gives a prediction on the prefactor in the relation of $\Phi_{rms}$ (and $E_{\Phi}$) apart from its linear dependence on $u_{rms}$ (and $E_{u}$). These predictions are tested experimentally in the next section.

\begin{figure}[htbp!]
	\centering
	\includegraphics[width=\linewidth]{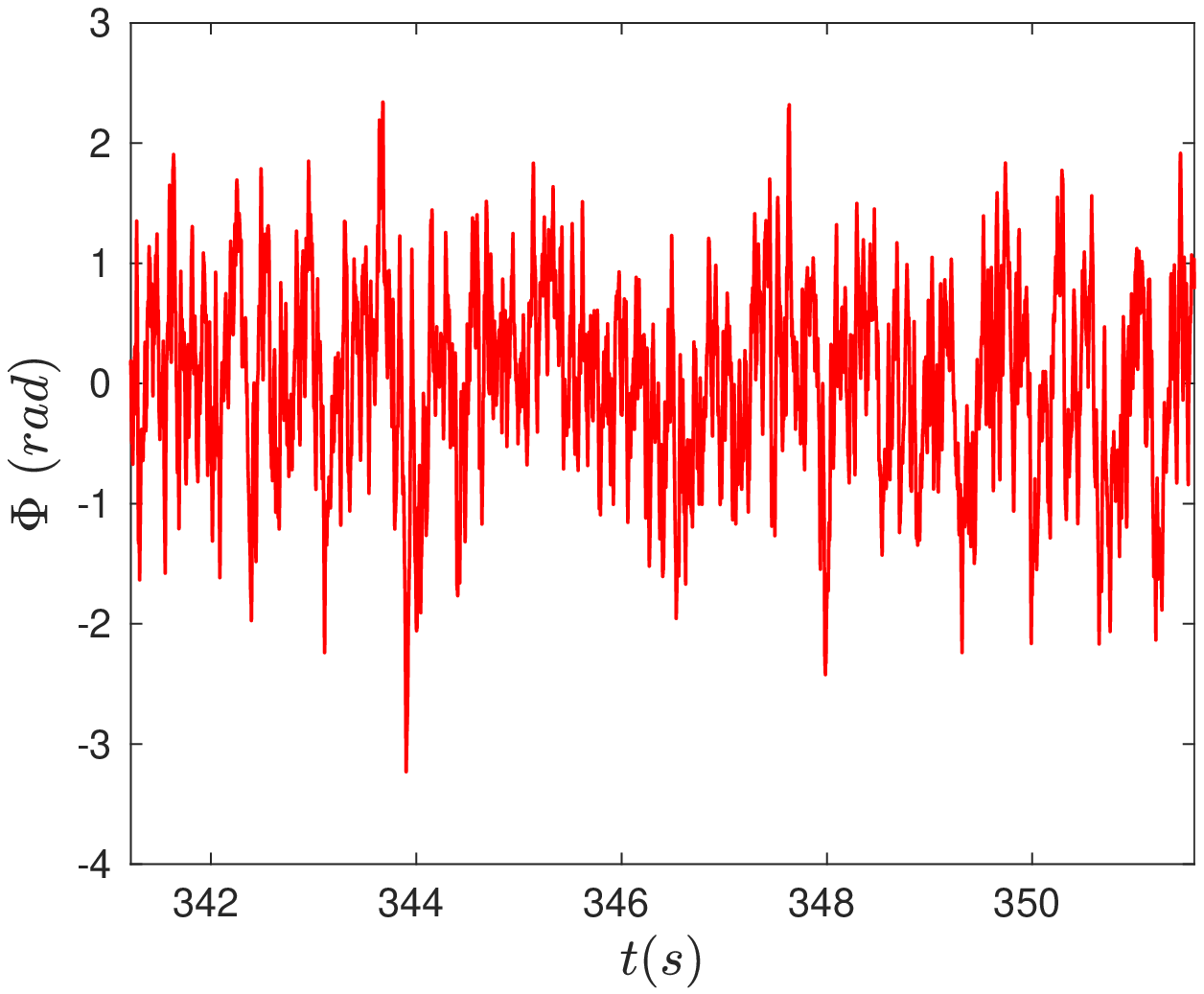}
	\includegraphics[width=\linewidth]{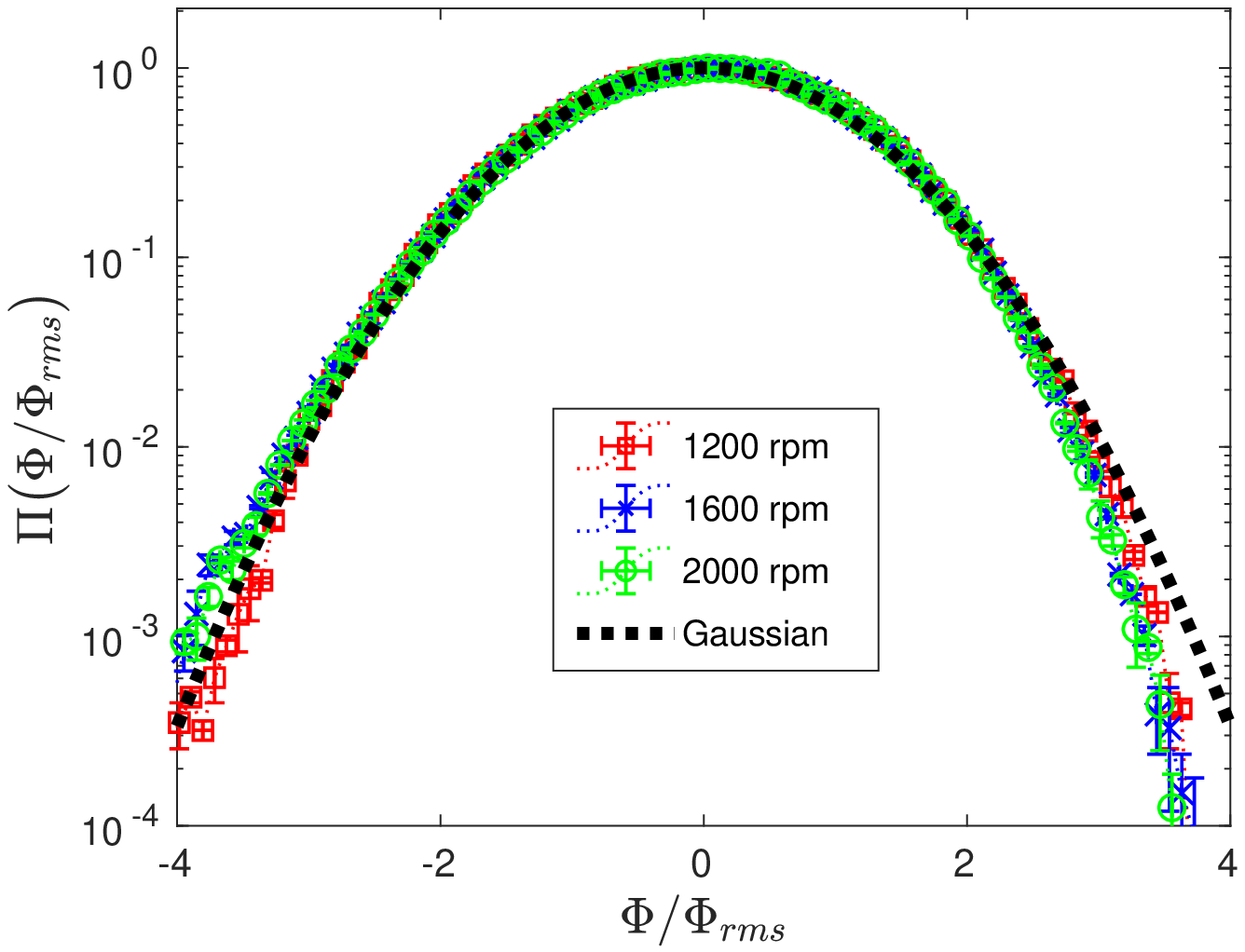}
	\includegraphics[width=\linewidth]{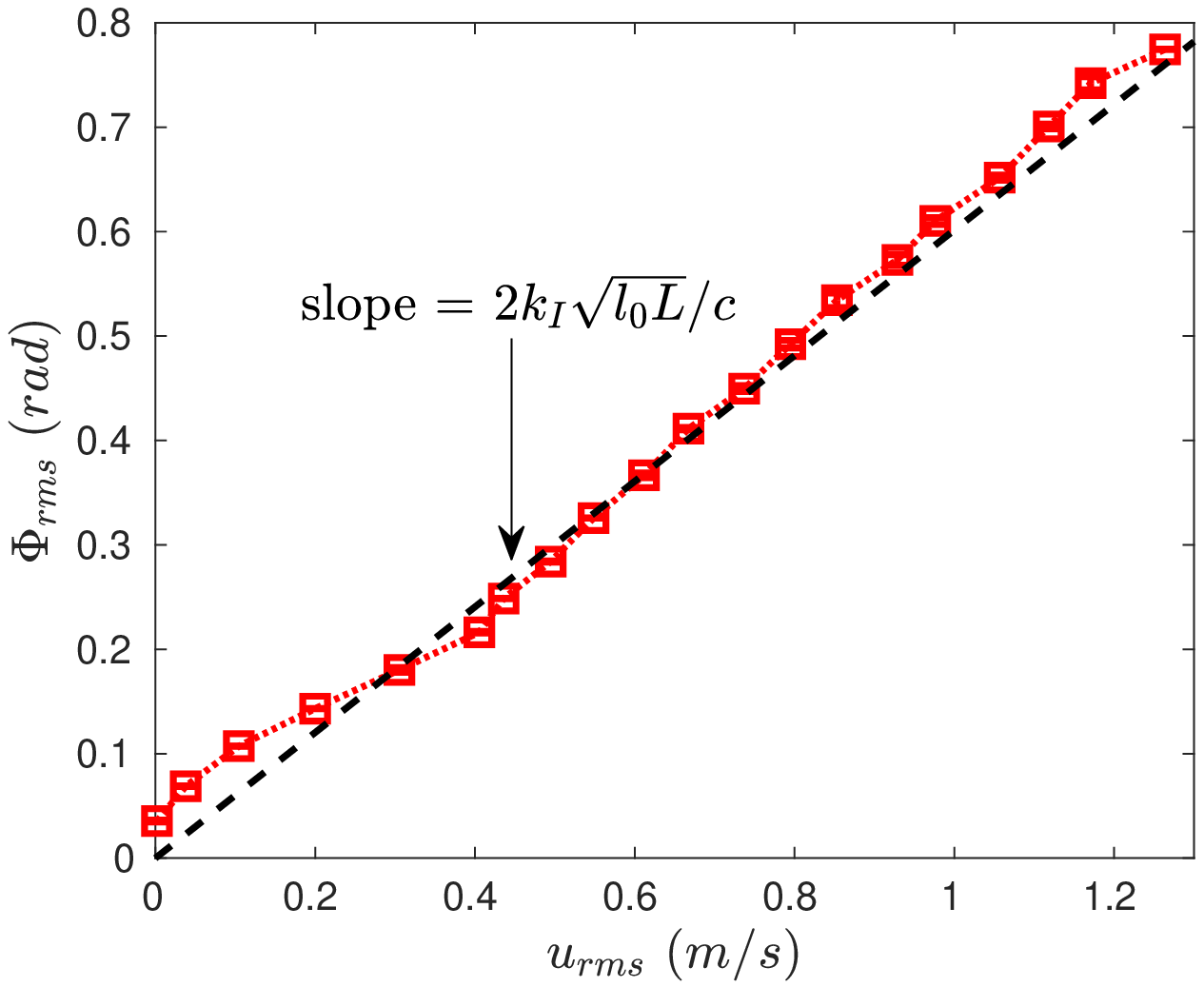}
	\caption{(a) Time series of phase fluctuations of the incident acoustic wave for $\Omega = 2000$ rpm. (b) PDFs of phase fluctuations normalized by their respective RMS values for $\Omega = 1200$ rpm ({\tiny$\textcolor{red}{\square}$}), $1600$ rpm ({\tiny$\textcolor{blue}{\square}$}) and $2000$ rpm ({\tiny$\textcolor{green}{\square}$}). Dashed line shows the Gaussian distribution. (c) The RMS of phase fluctuations $\Phi_{rms}$ vs the RMS of velocity fluctuations $u_{rms}$. Linear dependence is observed as predicted by eqn.~\ref{eqn:4}.}
	\label{fig:2}
\end{figure}

\section{Experimental setup and results} 

We generate a \vk swirling flow in air between two counter-rotating disks in a closed cylinder of diameter $130$ mm made out of copper of $2$ mm thickness (see fig.~\ref{fig:1}a). In order to maintain fixed the temperature at the boundaries, a copper tubing of outer diameter $10$ mm is welded to the cylindrical part of the container and a  water bath circulates water at a given temperature within $\pm 0.01$ K. 

We use two loop-controlled brushless DC motors rotating at an angular velocity $\Omega$ up to $2000$ rotations per minute (rpm). Each of the motor drives a disk with four curved blades. The thickness of the disks and the height of the blades are both $7.5$ mm. Holes drilled on the surface of the cylinder provide access to the probes; one 1D hot-wire velocity probe and two acoustic transducers all of which are placed in the midplane, with the acoustic transducers flushed to the wall (fig.~\ref{fig:1}b). The power dissipated per unit mass by the turbulent flow $\langle \epsilon \rangle$ is measured from the power required by the motors to maintain the flow, ranging from $0$ to $500$ m$^2$/s$^{3}$. This gives the Kolmogorov length scale $\eta = \big(\nu^{3}/\langle \epsilon \rangle\big)^{1/4} \geq 50$ $\mu$m with $\nu = 1.5 \times 10^{-5}$~m$^2$/s the kinematic viscosity of air at room temperature. We evaluate the Taylor microscale $\lambda = \big( 15 \nu u_{rms}^{2}/\langle \epsilon \rangle \big)^{1/2} \sim O(1)$ mm and the Taylor microscale based Reynolds number $Re_{\lambda} \sim O(10^{2})$ using standard estimates for homogeneous and isotropic turbulent flows~\cite{tennekes1972first}. The two point spatial correlation $\Gamma_{u}$ is evaluated using two 1D hot-wire probes moved radially about the center axis in the midplane. An exponential decay is observed with the characteristic length scale $l_{0} = 5 \ \text{mm}$ which does not vary with $\Omega$. 

The acoustic transducers (ITC-9073) have a diameter of $12$ mm and emit acoustic waves at frequency $f_{I} = 230$ kHz ($k_{I} = 4.13\times10^{3}$ m$^{-1}$). The transducers are surrounded by sound absorbing foam to absorb any reflected acoustic waves not in LOS propagation. A sine wave signal is supplied to the emitting transducer and a lock-in amplifier (Stanford Research Systems) analyzes  the signal measured by the receiving transducer. The lock-in amplifier directly measures the fluctuations in amplitude and phase of the incident wave as it propagates through the bulk of the turbulent flow from the emitter to the receiver. Figs.~\ref{fig:2}(a) and \ref{fig:2}(b) show the time series of phase fluctuations and its corresponding probability density function (PDF) denoted by $\Pi$, for $\Omega = 2000$ rpm. We observe that the PDFs are Gaussian for the range of rotation rates studied in this experiment. This is a consequence of the phase fluctuations being the integral of velocity fluctuations (see eqn. \ref{eqn:2}) which in turbulent flows are generically observed to follow a Gaussian distribution \cite{frisch1995turbulence}. Eqn.~\ref{eqn:4} predicts that the RMS of phase fluctuations $\Phi_{rms}$ would be proportional to $u_{rms}$. The constant of proportionality is a function of $k_{I},c,l_{0}$ and $L$, all of which are independent of $\Omega$ for the current experimental setup. As seen in fig.~\ref{fig:2}c, eqn.~\ref{eqn:4} correctly describes the behaviour of $\Phi_{rms}$ and does not involve any fitting parameter.

\begin{figure}[ht!]
	\centering
	\includegraphics[height=6cm,width=\linewidth]{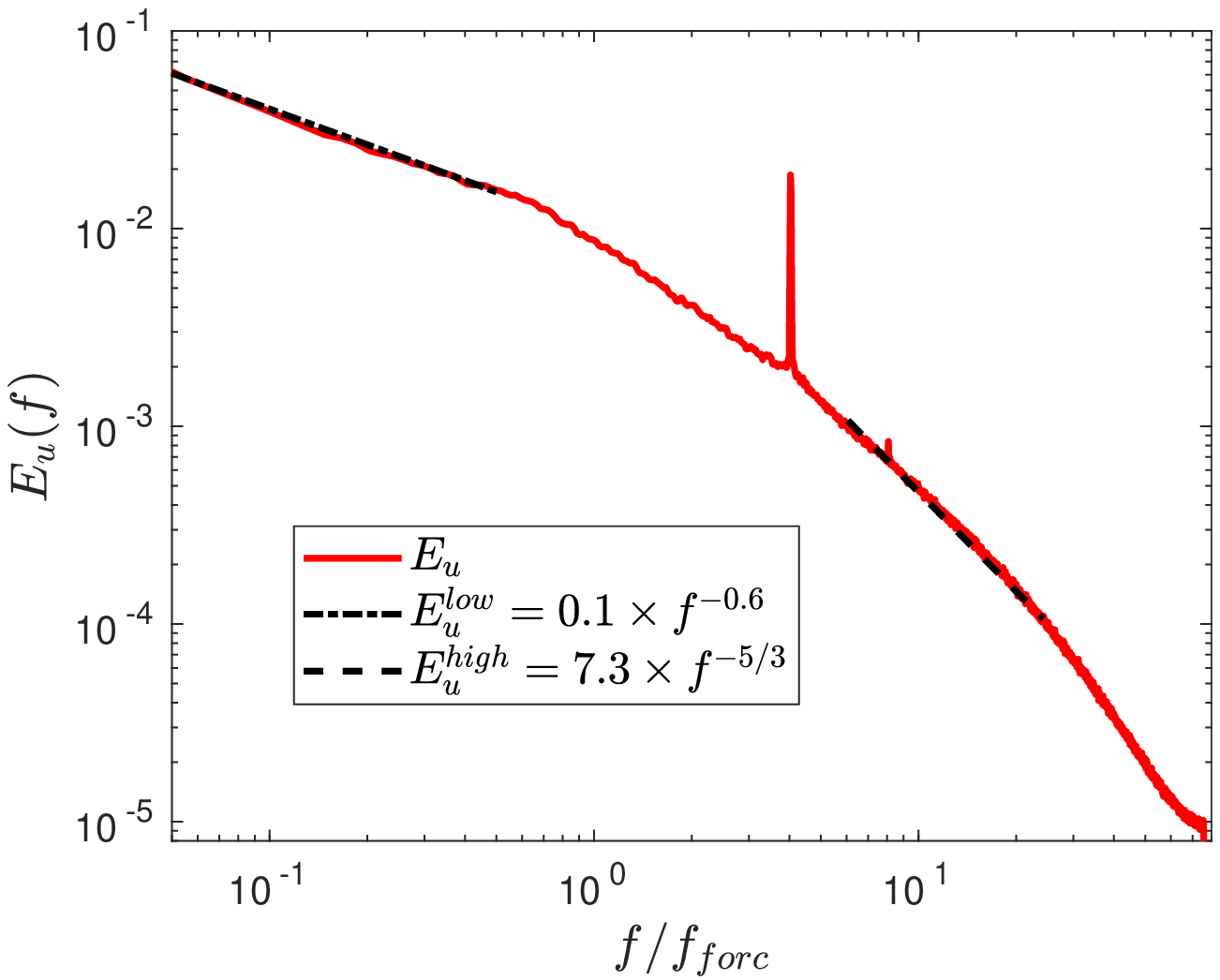}
	\includegraphics[height=6cm,width=\linewidth]{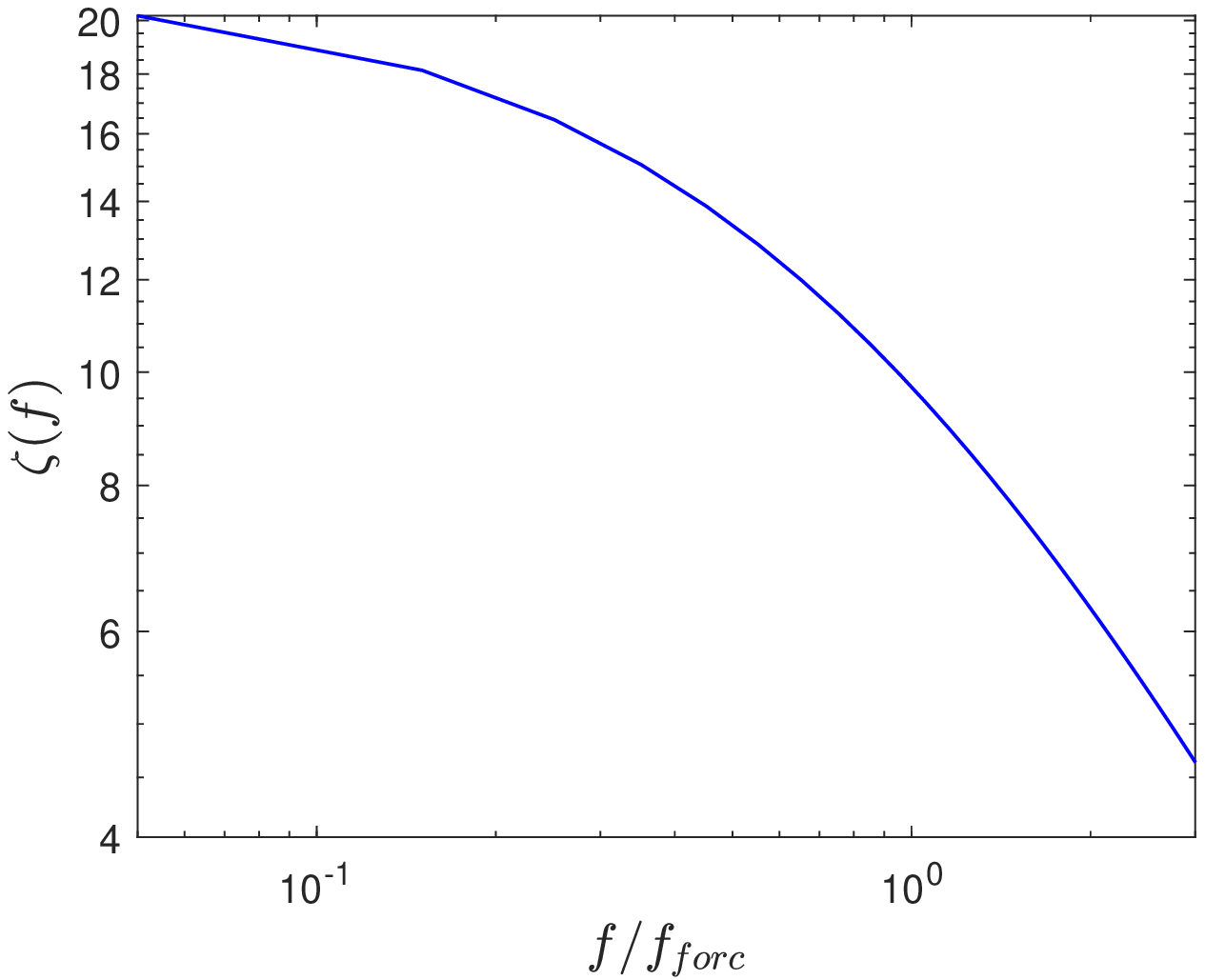}
	\includegraphics[height=6cm,width=\linewidth]{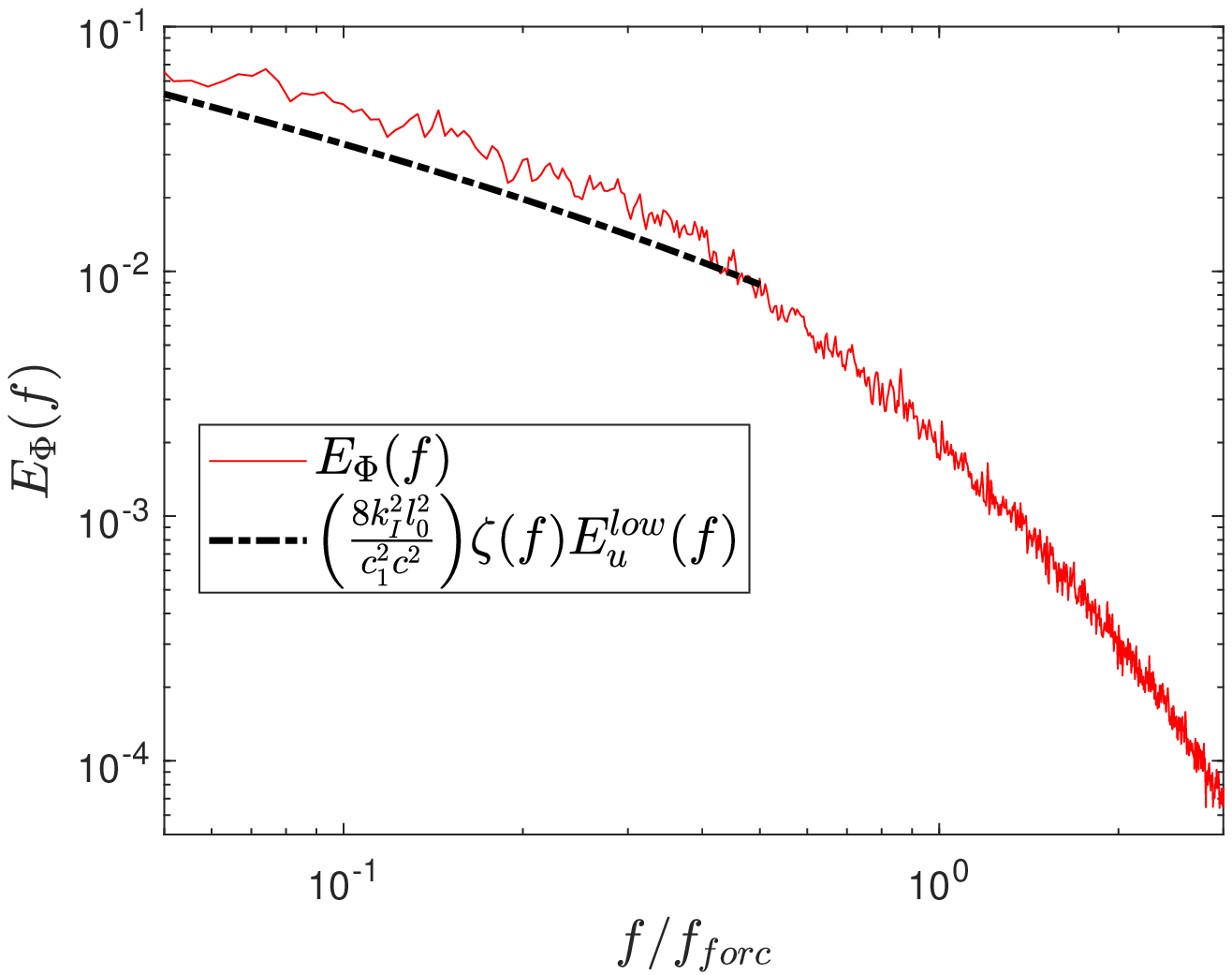}
	\caption{(a) Energy spectrum $E_{u}(f)$ obtained using 1D hot-wire probe. Two distinct power laws are observed with $E_{u}^{low} = 0.1 \times f^{-0.6}$ at low frequencies (dotted-dashed line) and $E_{u}^{high} = 7.3 \times f^{-5/3}$ at high frequencies (dashed line). (b) Function $\zeta(f)$ as given by eqn.~\ref{eqn:9} for $L/l_{0}=20$, $u_{rms} = 1.2$ m/s and $l_{0} = 5$ mm which correspond to the experimentally obtained values for $\Omega = 2000$ rpm. (c) Power spectrum of phase fluctuations $E_{\Phi}$. Dotted-dashed line denotes the prediction by eqn.~\ref{eqn:8} at low frequencies.}
	\label{fig:3}
\end{figure}

We now focus on the test of $E_{\Phi}(f)$ as predicted by eqn.~\ref{eqn:8}. Fig.~\ref{fig:3}(a) shows an example of PSD of velocity fluctuations obtained from the 1D hot-wire probe at a location as shown in fig.~\ref{fig:1} and for $\Omega = 2000$ rpm. The frequency is normalized with the forcing frequency $f_{forc} = \Omega/60$. We observe two power-laws, one at low frequency with $E_{u}^{low} = 0.1 \times f^{-0.6}$ and one at higher frequencies with $E_{u}^{high} = 7.3 \times f^{-5/3}$.
A similar low-frequency behaviour in PSD was also observed by Ravelet \cite{ravelet2005bifurcations} albeit with a slighltly different exponent who attributed it to coherent structures in the shear layer of \vk flow driven by disks with blades. The function $\zeta(f)$ is a monotonically decreasing function as seen in fig.~\ref{fig:3}(b) where we have taken the experimental values of the physical parameters. The theory relies on the limit of geometrical acoustics and is thus expected to be valid for large length scales of the velocity fluctuations, which corresponds to small frequencies in the temporal domain.
We thus use $E_{u}^{low}$  as measured for small frequencies and from eqn.~\ref{eqn:8} we obtain the prediction shown by dotted line in fig.~\ref{fig:3}(c). The slope is correctly predicted by the model and the measured $E_\Phi$ differs from the prediction only by a multiplicative factor of $1.3$.  The agreement is indeed satisfactory as several hypotheses for the model can be questioned, such as the homogeneity and isotropy of flow properties.

\section{Conclusion}

We have investigated the statistics of phase fluctuations of an acoustic wave travelling through a turbulent flow in line of sight (LOS) propagation. Building on the theory by Tatarskii \cite{tatarski2016wave} and on our previous experimental results \cite{prabhudesai2021coherence}, we obtain predictions for the RMS of phase fluctuations and for their PSD. The latter is  shown to be  related to the spatio-temporal  coherence of the velocity fluctuations. We have experimentally verified these predictions in a \vk swirling flow.

Acoustic measurements thus offer a non-intrusive way of measuring the integral length scale, and, under some conditions the coherence of the turbulent velocity fluctuations. The procedure is as follows: first the power dissipated per unit mass $\eavg{\epsilon}$ is estimated from the power consumed by the forcing driving the turbulent flow as was done for this experiment. The quantity $\eavg{\epsilon}$ has been demonstrated to follow the large-scale scaling $\eavg{\epsilon} = C_{\epsilon} u_{rms}^{3}/l_{0}$ with $C_{\epsilon} \approx 0.4$ \cite{Pope_2000,Vassilicos_2015}. Thus from the measurement of $\eavg{\epsilon}$ and $\Phi_{rms}$, and using the above relation along with eqn.~\ref{eqn:4}, one obtains the quantities $u_{rms}$ and $l_{0}$. We have used this method for the current experiment and verified that the values of these quantities thus obtained are in agreement with their measured values. 
	
To evaluate the coherence function amounts to evaluating the constants $c_{1}$ and $c_{2}$ defined in eqn.~\ref{eqn:7}. To do so non-intrusively using eqn.~\ref{eqn:8}, the functional form of $E_{u}(f)$ needs to be known. If we consider that $E_{u}(f)=\alpha f^{-p}$, then from $E_{\Phi}(f)$ obtained from acoustic measurements, the parameters $\alpha$, $p$, $c_{1}$ and $c_{2}$ can be evaluated using a fitting procedure. The number of fitting parameters is reduced if we have additional information on $E_{u}(f)$. This is for instance the case if the random sweeping hypothesis is valid \cite{Tennekes_1975,Chen_Kraichnan_1989}, which predicts the PSD for inertial scales in homogeneous, isotropic turbulence as $E_{u}(f) = \tilde{C} \eavg{\epsilon}^{2/3}u_{rms}^{2/3}f^{-5/3}$ with $\tilde{C} \approx 0.8$ \cite{Kit_Fernando_Brown_1995,Fung_Hunt_Malik_Perkins_1992,hunt1987big}. Then only $c_1$ and $c_2$ need to be evaluated from fitting. Note that the range of frequencies for which both the random sweeping hypothesis and geometrical acoustics would be applicable is given\footnote{ On scale by scale analysis of the turbulent flow, the theory of geometrical acoustics would be applicable for turbulent scales of length $l$ such that $\big( k_{I}l \big/ 2\pi\big) \gg 1$. In terms of frequency, it gives the condition $f \ll \big(u_{rms}\big/l_{0}\big)\big( l_{0}f_{I} \big/ c \big)^{2/3}$. On the other hand, the random sweeping effect would be observable for frequencies $f \gg \big( u_{rms} \big/ l_{0} \big)$. } by $\big(u_{rms}\big/l_{0}\big) \ll f \ll \big(u_{rms}\big/l_{0}\big)\big( l_{0}f_{I} \big/ c \big)^{2/3}$.

\acknowledgments
This work has been supported by the Agence nationale de la recherche (Grant No. ANR-17-CE30-0004), CEFIPRA (Project 6104-1), CNES (action 6291) and Labex ENS-ICFP.

\bibliographystyle{eplbib} 
\bibliography{biblio}

\end{document}